\documentclass[preprintnumbers,amsmath,amssymbm,prd]{revtex4}
\usepackage{epsfig}
\usepackage{graphicx}

\begin{document}
\title{Universality in the relaxation dynamics of the composed black-hole-charged-massive-scalar-field system:
The role of quantum Schwinger discharge}
\author{Shahar Hod}
\affiliation{The Ruppin Academic Center, Emeq Hefer 40250, Israel}
\affiliation{ }
\affiliation{The Hadassah Institute, Jerusalem 91010, Israel}
\date{\today}

\begin{abstract}
\ \ \ The quasinormal resonance spectrum
$\{\omega_n(\mu,q,M,Q)\}_{n=0}^{n=\infty}$ of charged massive scalar
fields in the charged Reissner-Nordstr\"om black-hole spacetime is
studied {\it analytically} in the large-coupling regime $qQ\gg M\mu$
(here $\{\mu, q\}$ are respectively the mass and charge coupling
constant of the field, and $\{M,Q\}$ are respectively the mass and
electric charge of the black hole). This physical system provides a
striking illustration for the validity of the universal relaxation
bound $\tau \times T \geq \hbar/\pi$ in black-hole physics (here
$\tau\equiv 1/\Im\omega_0$ is the characteristic relaxation time of
the composed black-hole-scalar-field system, and $T$ is the
Bekenstein-Hawking temperature of the black hole). In particular, it
is shown that the relaxation dynamics of charged massive scalar
fields in the charged Reissner-Nordstr\"om black-hole spacetime may
{\it saturate} this quantum time-times-temperature inequality.
Interestingly, we prove that potential violations of the bound by
light scalar fields are excluded by the Schwinger-type
pair-production mechanism (a vacuum polarization effect), a {\it
quantum} phenomenon which restricts the physical parameters of the
composed black-hole-charged-field system to the regime $qQ\ll
M^2\mu^2/\hbar$.
\end{abstract}
\bigskip
\maketitle

\section{Introduction}

Wheeler's celebrated conjecture that black holes are bald
\cite{Whee,Car} has played a key role in black-hole physics over the
last four decades \cite{Bek1,Chas,BekVec,Hart,Nun,Hod11}. This
`no-hair' conjecture asserts that static matter fields cannot be
supported in the exterior regions of asymptotically flat black-hole
spacetimes \cite{Whee,Car,Bek1,Chas,BekVec,Hart,Nun,Hod11}.

The no-hair conjecture \cite{Whee,Car} thus suggests a simple and
universal picture in which matter fields that propagate in the
exterior spacetime regions of newly born black holes would
eventually be scattered away to infinity or be absorbed into the
central black hole. This suggested scenario is supported by the fact
\cite{Vis} that the dynamics of fundamental fields in black-hole
spacetimes are characterized by ringdown ({\it damped}) oscillations
of the form $e^{-i\omega t}$ \cite{Notescc,Hodrc,HerR}. These
decaying oscillations, which are known as the quasinormal modes of
the composed black-hole-field system, are characterized by an
infinite spectrum of complex resonant frequencies
$\{\omega_n\}_{n=0}^{n=\infty}$ \cite{QNMs}.

In accord with the spirit of the no-hair conjecture, these decaying
modes characterize the relaxation dynamics of composed
black-hole-fundamental-field systems. In particular, the damped
quasinormal oscillations reflect the gradual decay of the matter
fields (the `hair') in the exterior spacetime regions of newly born
black holes \cite{Notetails,Tails}.

It should be emphasized, however, that existing no-hair theorems
\cite{Whee,Car,Bek1,Chas,BekVec,Hart,Nun,Hod11} say nothing about
the {\it timescale} it takes for a newly born black hole to become
bald. This characteristic relaxation timescale,
$\tau_{\text{relax}}$, is determined by the fundamental (least
damped) black-hole quasinormal resonance:
\begin{equation}\label{Eq1}
\tau_{\text{relax}}\equiv 1/\Im\omega_0\  .
\end{equation}
An interesting question naturally arises: How short can the
relaxation time (\ref{Eq1}) be?

A remarkably simple answer to this fundamental question was
suggested in \cite{HodTTT}: using standard ideas from information
theory and thermodynamics, it was argued in \cite{HodTTT} that the
characteristic relaxation time of a perturbed thermodynamic system
should be bounded from below by the time-times-temperature (TTT)
inequality \cite{Notetl}
\begin{equation}\label{Eq2}
\tau_{\text{relax}}\times T\geq{{\hbar}\over{\pi k_{\text{B}}}}\  ,
\end{equation}
where $T$ is the system temperature \cite{Noteunit}. In particular,
in the context of black-hole perturbation theory, it was argued in
\cite{HodTTT} that the imaginary part of the fundamental black-hole
quasinormal resonance should be bounded from above according to the
simple relation [see Eqs. (\ref{Eq1}) and (\ref{Eq2})]
\begin{equation}\label{Eq3}
\Im\omega_0\leq\pi T_{\text{BH}}\  ,
\end{equation}
where $T_{\text{BH}}$ is the black-hole temperature.

Interestingly, it was demonstrated numerically in \cite{HodTTT} that
the quasinormal resonances of spinning Kerr black holes indeed
conform to the suggested upper bound (\ref{Eq3}). Moreover, it was
proved analytically \cite{Hodan} (see also \cite{Als}) that
rapidly-rotating (near-extremal) Kerr black holes may actually {\it
saturate} the bound. Namely, their fundamental resonances are
characterized by the simple asymptotic relation \cite{Notelong}
\begin{equation}\label{Eq4}
\Im\omega_0\to\pi T_{\text{BH}}\ \ \ \text{as}\ \ \
T_{\text{BH}}\to0\ .
\end{equation}

\section{Charged scalar fields and Schwinger-type pair production}

The relaxation dynamics of charged matter fields in a newly born
{\it charged} Reissner-Nordstr\"om (RN) black-hole spacetime
\cite{Notenhc,Hodth} was studied analytically in
\cite{Hodch1,Hodch2}. In particular, it was shown that the
fundamental quasinormal resonances of {\it massless} ($\mu=0$)
charged scalar fields in the charged black-hole spacetime are
characterized by the simple relation \cite{Hodch2}
\begin{equation}\label{Eq5}
\Im\omega_0=\pi T_{\text{BH}}\big[1+O(qQ/l^2)\big]\ \ \ \text{for}\
\ \ \{\mu=0\ \ \text{and}\ \ l\ll qQ\ll l^2\}\ ,
\end{equation}
where $T_{\text{BH}}$ is the RN black-hole temperature [see Eq.
(\ref{Eq42}) below]. Here $q,l$, and $Q$ are respectively the charge
coupling constant of the field, the spherical harmonic index of the
field mode [see Eq. (\ref{Eq14}) below], and the electric charge of
the black hole \cite{NoteQ}. A similar result was later obtained in
\cite{Konf} for charged massless scalar fields in the regime $qQ\gg
(l+1)^2$. Most recently, Ref. \cite{Ric} extended this result to
higher orders in the small quantity $(l+1)^2/qQ$, finding
\begin{equation}\label{Eq6}
\Im\omega_0=\pi T_{\text{BH}}\Big(1+\pi
T_{\text{BH}}{{r_+-3r_-}\over{4q^2Q^2}}\Big)\ \ \ \text{for}\ \ \
\{\mu=0\ \ \text{and}\ \ qQ\gg (l+1)^2\}\  ,
\end{equation}
where $r_{\pm}=M\pm(M^2-Q^2)^{1/2}$ are the horizon radii of the RN
black hole (here $M$ is the black-hole mass).

Interestingly, the result (\ref{Eq6}) suggests a violation of the
universal relaxation bound (\ref{Eq2}) [or equivalently, a violation
of the black-hole quasinormal bound (\ref{Eq3})] for charged RN
black holes in the regime \cite{Noterg}
\begin{equation}\label{Eq7}
Q/M<{{\sqrt{3}}\over{2}}\  .
\end{equation}
In particular, {\it if} the gravitational collapse of a charged {\it
massless} scalar field with $qQ\gg (l+1)^2$ could end up in the
formation of a charged RN black hole with $Q/M<{{\sqrt{3}}/{2}}$,
then the characteristic relaxation time [see Eqs. (\ref{Eq1}) and
(\ref{Eq6})]
\begin{equation}\label{Eq8}
\tau_{\text{relax}}={{1}\over{\pi T_{\text{BH}}}}\times\Big(1-\pi
T_{\text{BH}}{{r_+-3r_-}\over{4q^2Q^2}}\Big)<{{1}\over{\pi
T_{\text{BH}}}}\ \ \ \text{for}\ \ \ \{\mu=0, \ \ \text{and}\ \
qQ\gg (l+1)^2\}\
\end{equation}
of the newly born charged black hole would violate the universal
relaxation bound (\ref{Eq2}) \cite{NoteQM}.

One naturally wonders: Is there a physical mechanism which can
prevent the apparent violation (\ref{Eq8}) of the universal
relaxation bound? It seems that no classical effect can prevent this
violation. However, as we shall show below, the Schwinger-type
pair-production mechanism \cite{Schw1,Schw2,Schw3,Schw4}, a purely
{\it quantum} effect, ensures the validity of the ({\it quantum})
universal relaxation bound (\ref{Eq2}).

In particular, vacuum polarization effects set an upper bound on the
electric field strength of charged RN black holes
\cite{Schw1,Schw2,Schw3,Schw4}: $E_+<E_{\text{c}}\equiv
\mu^2/q\hbar$, where $E_+=Q/r^2_+$ is the electric field at the
black-hole horizon. In fact, it was shown in \cite{Schw4} that the
Schwinger-type pair production mechanism becomes effective (that is,
with pair production probability of almost $100\%$) already at
$E\simeq 0.03E_{\text{c}}\ll E_{\text{c}}$. Thus, the quantum
Schwinger discharge of charged RN black holes (the vacuum
polarization effect) sets an upper bound on the electric field
strengths of the black holes: $Q/r^2_+\ll \mu^2/q\hbar$, or
equivalently
\begin{equation}\label{Eq9}
qQ\ll\mu^2r^2_+\  .
\end{equation}
The inequality (\ref{Eq9}) implies, in particular, that the charged
{\it massless} scalar fields studied in \cite{Hodch2,Konf,Ric} are
physically unacceptable in the {\it quantum} regime.

In particular, taking cognizance of the fact that the suggested
universal relaxation bound (\ref{Eq2}) is intrinsically a {\it
quantum} phenomenon \cite{Notehb}, one realizes that a
self-consistent test of the bound validity in the context of charged
gravitational collapse must include {\it massive} charged fields
which respect the {\it quantum} \cite{Notebq} inequality
(\ref{Eq9}).

The main goal of the present paper is thus to test the bound
validity in a self-consistent manner, that is, in the physically
acceptable regime (\ref{Eq9}). To that end, we shall study below the
relaxation dynamics of a newly born charged RN black hole. In
particular, we shall analyze the quasinormal resonance spectrum (the
characteristic {\it damped} oscillations) of charged {\it massive}
perturbations fields [see Eq. (\ref{Eq9})] in the background of the
newly born charged black hole.

\section{Description of the system}

We analyze the dynamics of a charged massive scalar field linearly
coupled to a Reissner-Nordstr\"om black hole of mass $M$ and
electric charged $Q$. The charged RN black-hole spacetime is
described by the line element \cite{Chan}
\begin{equation}\label{Eq10}
ds^2=-f(r)dt^2+{1\over{f(r)}}dr^2+r^2(d\theta^2+\sin^2\theta
d\phi^2)\ ,
\end{equation}
where
\begin{equation}\label{Eq11}
f(r)=1-{{2M}\over{r}}+{{Q^2}\over{r^2}}\  .
\end{equation}
The radii of the black-hole (event and inner) horizons,
\begin{equation}\label{Eq12}
r_{\pm}=M\pm(M^2-Q^2)^{1/2}\  ,
\end{equation}
are determined by the zeros of $f(r)$.

The dynamics of a scalar field $\Psi$ of mass $\mu$ and charge
coupling constant $q$ \cite{Noteqm} in the black-hole spacetime is
governed by the Klein-Gordon wave equation
\cite{HodPirpam,Stro,HodCQG2}
\begin{equation}\label{Eq13}
[(\nabla^\nu-iqA^\nu)(\nabla_{\nu}-iqA_{\nu})-\mu^2]\Psi=0\  ,
\end{equation}
where $A_{\nu}=-\delta_{\nu}^{0}{Q/r}$ is the electromagnetic
potential of the charged RN black hole.

It is convenient to decompose the scalar field $\Psi$ in the form
\begin{equation}\label{Eq14}
\Psi(t,r,\theta,\phi)=\int\Sigma_{lm}e^{im\phi}S_{lm}(\theta)R_{lm}(r;\omega)e^{-i\omega
t} d\omega\ ,
\end{equation}
where the integers $l$ and $m$ are the spherical harmonic index and
the azimuthal harmonic index of the field mode, respectively
\cite{Noteom}. Substituting the field decomposition (\ref{Eq14})
into the Klein-Gordon wave equation (\ref{Eq13}), one finds
\cite{HodPirpam,Stro,HodCQG2} that the radial and angular functions
[$R(r)$ and $S(\theta)$, respectively] are determined by two
differential equations of the confluent Heun type \cite{Heun,Abram}.
These two equations are coupled by the angular eigenvalues
(separation constants) $K_l=l(l+1)$ with $l\geq |m|$. The radial
wave equation is given by \cite{HodPirpam,Stro,HodCQG2}
\begin{equation}\label{Eq15}
\Delta{{d} \over{dr}}\Big(\Delta{{dR}\over{dr}}\Big)+UR=0\ ,
\end{equation}
where $\Delta=r^2f(r)$, and
\begin{equation}\label{Eq16}
U=(\omega r^2-qQr)^2 -\Delta[\mu^2r^2+l(l+1)]\  .
\end{equation}

Defining the new radial function
\begin{equation}\label{Eq17}
\psi=rR\  ,
\end{equation}
and using the ``tortoise" radial coordinate $y$, which is defined by
the relation
\begin{equation}\label{Eq18}
dy={{dr}\over{f(r)}}\  ,
\end{equation}
one can write the radial equation (\ref{Eq15}) in the form of a
Schr\"odinger-like wave equation
\begin{equation}\label{Eq19}
{{d^2\psi}\over{dy^2}}+V\psi=0\  .
\end{equation}
Here
\begin{equation}\label{Eq20}
V=V(r;M,Q,\omega,q,\mu,l)=\Big(\omega-{{qQ}\over{r}}\Big)^2-{{f(r)H(r)}\over{r^2}}\
\end{equation}
plays the role of an effective radial potential, where
\begin{equation}\label{Eq21}
H(r;M,Q,\mu,l)= \mu^2r^2+l(l+1)+{{2M}\over{r}}-{{2Q^2}\over{r^2}}\ .
\end{equation}

The Schr\"odinger-like radial wave equation (\ref{Eq19}) is
supplemented by the physically motivated boundary conditions of
purely ingoing waves at the black-hole horizon and purely outgoing
waves at spatial infinity \cite{Detw}. That is,
\begin{equation}\label{Eq22}
\psi \sim
\begin{cases}
e^{-i (\omega-qQ/r_+)y} & \text{ as\ \ \ } r\rightarrow r_+\ \
(y\rightarrow -\infty)\ ; \\ y^{-iqQ}e^{i\sqrt{\omega^2-\mu^2} y} &
\text{ as\ \ \ } r\rightarrow\infty\ \ (y\rightarrow \infty)\  .
\end{cases}
\end{equation}
These boundary conditions single out a {\it discrete} spectrum of
complex eigenvalues $\{\omega_n(M,Q,\mu,q,l)\}_{n=0}^{n=\infty}$
which correspond to the quasinormal resonances of the charged
massive scalar field in the charged RN black-hole spacetime. The
main goal of the present paper is to determine {\it analytically}
these characteristic resonances of the composed
black-hole-scalar-field system in the physically acceptable regime
(\ref{Eq9}).

\section{The quasinormal resonance spectrum of the composed black-hole-charged-field system}

In the present section we shall analyze the resonance spectrum of
the charged massive scalar fields in the charged RN black-hole
spacetime. We shall focus on the large-coupling regime
\begin{equation}\label{Eq23}
qQ\gg\text{max}\{\mu r_+, l+1\}\  .
\end{equation}
As we shall show below, in the regime (\ref{Eq23}) the radial
potential (\ref{Eq20}) has the form of an effective potential
barrier whose fundamental scattering resonances can be studied
analytically using standard WKB methods \cite{WKB1,WKB2,WKB3,Will}.

In particular, in the large-coupling regime (\ref{Eq23}) the maximum
$r_0$ of the potential barrier (\ref{Eq20}) is located in the
vicinity of the black-hole horizon:
\begin{equation}\label{Eq24}
{{r_0-r_+}\over{r_+-r_-}}\ll1\  .
\end{equation}
To see this, it proves useful to introduce the dimensionless
variables
\begin{equation}\label{Eq25}
x= {{r-r_+}\over{r}}\ \ \ ; \ \ \ \tau={{r_+-r_-}\over{r_+}}\ \ \ ;
\ \ \ \varpi= {{\omega r_+}\over{qQ}}-1\ ,
\end{equation}
in terms of which the effective radial potential (\ref{Eq20})
becomes
\begin{equation}\label{Eq26}
V(x;\varpi)=\Big({{qQ}\over{r_+}}\Big)^2(x+\varpi)^2-{{H(r_+)\tau}\over{r^2_+}}x
[1+O(x/\tau)]\  .
\end{equation}
Differentiating (\ref{Eq26}) with respect to the dimensionless
radial coordinate $x$, one finds
\begin{equation}\label{Eq27}
x_0+\varpi={{H(r_+)\tau}\over{2q^2Q^2}}\ll 1\
\end{equation}
for the location $x_0$ of the maximum of the radial potential
barrier \cite{Noteleq}.

As shown in \cite{WKB1,WKB2}, the characteristic WKB resonance
equation for the scattering resonances of the Schr\"odinger-like
wave equation (\ref{Eq19}) can be written in the form
\begin{equation}\label{Eq28}
iK=n+{1\over 2}+\Lambda(n)+O[\Omega(n)]\ ,
\end{equation}
where \cite{WKB2}
\begin{equation}\label{Eq29}
K={{V_0}\over{\sqrt{2V^{(2)}_0}}}\  ,
\end{equation}
\begin{equation}\label{Eq30}
\Lambda(n)={{1}\over{\sqrt{2V^{(2)}_0}}}\Big[{{1+(2n+1)^2}\over{32}}\cdot{{V^{(4)}_0}\over{V^{(2)}_0}}-
{{28+60(2n+1)^2}\over{1152}}\cdot\Big({{V^{(3)}_0}\over{V^{(2)}_0}}\Big)^2\Big]\
,
\end{equation}
and the cumbersome expression for the sub-leading correction term
$\Omega(n)$ is given by Eq. (1.5b) of \cite{WKB2}. Here
$V^{(k)}_0\equiv d^{k}V/dy^{k}$ are evaluated at the maximum $y=y_0$
of the effective potential barrier $V(y)$.

Taking cognizance of Eqs. (\ref{Eq26}), (\ref{Eq27}), (\ref{Eq29}),
and (\ref{Eq30}), one finds
\begin{equation}\label{Eq31}
K={{H(r_+)}\over{2qQ}}\cdot{{\varpi-{{H(r_+)\tau}\over{4q^2Q^2}}}
\over{ {{H(r_+)\tau}\over{2q^2Q^2}}-\varpi}}\  ,
\end{equation}
\begin{equation}\label{Eq32}
\Lambda(n)=-{{2qQ}\over{H(r_+)}}\cdot (n+{1\over 2})^2\  ,
\end{equation}
and
\begin{equation}\label{Eq33}
\Omega(n)=O[(qQ/H(r_+))^2(n+1/2)^3]\  .
\end{equation}
Note that the quantum constraint (\ref{Eq9}) implies that, in the
regime $n\ll H(r_+)/qQ$ \cite{Noteie1}, the various terms that
appear on the r.h.s of the WKB resonance equation (\ref{Eq28}) are
characterized by the strong inequalities \cite{Noteie1}
\begin{equation}\label{Eq34}
{{\Lambda(n)}\over{n+{1\over 2}}}=O[(n+1/2)qQ/H(r_+)]\ll1\ \ \
\text{and}\ \ \
{{\Omega(n)}\over{\Lambda(n)}}=O[(n+1/2)qQ/H(r_+)]\ll1\  .
\end{equation}

Substituting (\ref{Eq31})-(\ref{Eq33}) into the WKB resonance
equation (\ref{Eq28}), one finds
\begin{equation}\label{Eq35}
\varpi_R={{H(r_+)\tau}\over{4q^2Q^2}}\cdot
{{H^2(r_+)+2(2n+1)^2q^2Q^2[1-(2n+1)qQ/H(r_+)]^2}\over{H^2(r_+)+(2n+1)^2q^2Q^2[1-(2n+1)qQ/H(r_+)]^2}}\cdot
\big\{1+O[(qQ/H(r_+))^2]\big\}
\end{equation}
and
\begin{equation}\label{Eq36}
\varpi_I=-i{{H(r_+)\tau}\over{4q^2Q^2}}\cdot
{{(2n+1)qQH(r_+)[1-(2n+1)qQ/H(r_+)]}\over{H^2(r_+)+(2n+1)^2q^2Q^2[1-(2n+1)qQ/H(r_+)]^2}}\cdot
\big\{1+O[(qQ/H(r_+))^2]\big\}\  .
\end{equation}
Using the strong inequality \cite{Noteie1}
\begin{equation}\label{Eq37}
{{qQ}\over{H(r_+)}}\ll1\  ,
\end{equation}
one can write the expressions (\ref{Eq35})-(\ref{Eq36}) in the form
\begin{equation}\label{Eq38}
\varpi_R={{H(r_+)\tau}\over{4q^2Q^2}}\cdot
\big\{1+O[(qQ/H(r_+))^2]\big\}
\end{equation}
and
\begin{equation}\label{Eq39}
\varpi_I=-i{{\tau}\over{2qQ}}(n+1/2)[1-(2n+1)qQ/H(r_+)]\cdot
\big\{1+O[(qQ/H(r_+))^2]\big\}\ .
\end{equation}

Finally, taking cognizance of Eq. (\ref{Eq25}), one finds
\begin{equation}\label{Eq40}
\omega_n={{qQ}\over{r_+}}+{{H(r_+)\tau}\over{4qQr_+}}-i{{\tau}\over{2r_+}}\cdot
\Big[n+{1\over
2}-{{2qQ}\over{H(r_+)}}\cdot\Big(n+{{1}\over{2}}\Big)^2\Big]\ \ \ ;
\ \ \ n=0,1,2,...
\end{equation}
for the characteristic quasinormal resonances of the composed
black-hole-charged-massive-scalar-field system. It is worth
emphasizing again that the resonance spectrum (\ref{Eq40}) is valid
in the physically acceptable regime (\ref{Eq9}).

\section{Summary and Discussion}

We have analyzed the quasinormal resonance spectrum of charged
massive scalar fields in the charged Reissner-Nordstr\"om black-hole
spacetime. Our main goal in this paper was to test the validity of
the suggested universal relaxation bound (\ref{Eq2}) in the context
of black-hole physics (and, in particular, in the context of
dynamical `hair' shedding of newly born charged RN black holes).

It was first pointed out that charged {\it massless} fields in the
regime $qQ\gg (l+1)^2$ can violate the relaxation bound (\ref{Eq2}).
In particular, we have stressed the fact that, {\it if} the
gravitational collapse of a charged massless scalar field in the
large coupling regime $qQ\gg (l+1)^2$ could end up in the formation
of a RN black hole with $Q/M<{{\sqrt{3}}/{2}}$, then the
characteristic relaxation time (\ref{Eq8}) of the newly born charged
black hole would violate the suggested universal relaxation bound
(\ref{Eq2}).

However, given the fact that the universal relaxation bound
(\ref{Eq2}) is intrinsically a {\it quantum} phenomenon
\cite{Notehb}, we have argued that a self-consistent test of the
bound in the context of charged gravitational collapse must include
{\it massive} charged fields which respect the {\it quantum}
\cite{Notebq} inequality (\ref{Eq9}). This inequality is a direct
consequence of the Schwinger discharge mechanism (a vacuum
polarization effect), a quantum phenomenon which sets a bound on the
physically allowed parameters of the composed
black-hole-charged-massive-scalar-field system:
$qQ\ll\mu^2r^2_+/\hbar$.

We have shown that the characteristic quasinormal resonances of the
composed black-hole-charged-field system can be studied {\it
analytically} in the regime
\begin{equation}\label{Eq41}
\text{max}\{\mu r_+, l+1\}\ll qQ\ll \mu^2 r^2_+\  .
\end{equation}
[The left inequality in (\ref{Eq41}) corresponds to the large
coupling assumption (\ref{Eq23}), whereas the right inequality in
(\ref{Eq41}) corresponds to the quantum constraint (\ref{Eq9})
imposed by the Schwinger-type pair production mechanism]. In
particular, we have derived the resonance spectrum (\ref{Eq40}) in
the regime (\ref{Eq41}).

Interestingly, as we shall now show, the quasinormal resonance
spectrum (\ref{Eq40}) can be expressed in terms of the physical
parameters of the composed black-hole-charged-field system: the
black-hole temperature $T_{\text{BH}}$, the black-hole electrostatic
potential $\Phi_+$, the black-hole electric field strength $E_+$,
and the critical electric field $E_{\text{c}}$ for Schwinger-type
pair production of the charged massive particles. In particular,
using the relations
\begin{equation}\label{Eq42}
T_{\text{BH}}={{r_+-r_-}\over{4\pi r^2_+}}\ \ \ , \ \ \
\Phi_+={{Q}\over{r_+}}\ \ \ , \ \ \ E_+={{Q}\over{r^2_+}}\ \ \ , \ \
\ E_{\text{c}}={{\mu^2}\over{q}}\  ,
\end{equation}
one can write the resonance spectrum (\ref{Eq40}) in the form
\cite{Notemqq}:
\begin{equation}\label{Eq43}
\omega_n={q\Phi_+}+\pi T_{\text{BH}}{{E_{\text{c}}}\over{E_+}}-i2\pi
T_{\text{BH}}\cdot \Big[n+{1\over
2}-2{{E_+}\over{E_{\text{c}}}}\cdot \Big(n+{{1}\over{2}}\Big)^2
\Big]\ \ \ ; \ \ \ n=0,1,2,...\  .
\end{equation}

Finally, we note that the quasinormal spectrum (\ref{Eq43}) implies
the simple relation
\begin{equation}\label{Eq44}
\Im\omega_0=\pi T_{\text{BH}}\Big(1-{{E_+}\over{E_{\text{c}}}}\Big)
\end{equation}
for the fundamental (least damped) quasinormal resonance of the
composed black-hole-charged-field system. This relation yields
\cite{Notecc}
\begin{equation}\label{Eq45}
\tau_{\text{relax}}={{\hbar}\over{\pi
T_{\text{BH}}}}\Big(1+{{E_+}\over{E_{\text{c}}}}\Big)>{{\hbar}\over{\pi
T_{\text{BH}}}}\
\end{equation}
for the characteristic relaxation time of the newly born charged
black hole. This result [and, in particular, the last inequality in
(\ref{Eq45})] is in perfect harmony with the suggested universal
relaxation bound (\ref{Eq2}).

Moreover, the relation (\ref{Eq45}) reflects the fact that, to
leading order in the dimensionless small quantity
${{E_+}/{E_{\text{c}}}}$ \cite{Notecc}, the characteristic
relaxation time $\tau_{\text{relax}}={{\hbar}/{\pi
T_{\text{BH}}}}[1+O(E_+/E_{\text{c}})]$ of the composed
black-hole-field system in the physically acceptable regime
(\ref{Eq41}) is {\it universal}, that is, independent of the
charged-field parameters.

\bigskip
\noindent
{\bf ACKNOWLEDGMENTS}
\bigskip

This research is supported by the Carmel Science Foundation. I would
like to thank Yael Oren, Arbel M. Ongo, Ayelet B. Lata, and Alona B.
Tea for helpful discussions.



\begin{thebibliography}{99}
\bibitem{Notetails} At asymptotically late times, these exponentially decaying resonances are
followed by inverse power-law wave tails \cite{Tails}. These
late-time decaying tails are a consequence of the backscattering of
the wave fields from asymptotically far regions.
\bibitem{Notetl} Note that this bound can be regarded as a quantitative formulation of the third law of
thermodynamics.
\bibitem{Noteunit} We shall henceforth use natural units in which
$G=c=\hbar=k_{\text{B}}=1$.
\bibitem{Notelong} Note that the relation (\ref{Eq4}) implies extremely long relaxation times,
$\tau_{\text{relax}}\to 1/{\pi T_{\text{BH}}}$, for newly born
rapidly rotating ($T_{\text{BH}}\to0$) Kerr black holes.
\bibitem{Notenhc} It is worth emphasizing that
the no-hair theorem \cite{Bek1} has ruled out the existence of
static charged hairy black-hole-scalar-field configurations.
Moreover, recent studies \cite{Hodth} have shown that charged
Reissner-Nordstr\"om black holes cannot support stationary (or
exponentially growing in time) charged scalar configurations
(linearized scalar `clouds') in their exterior regions. It is
therefore expected that, during a charged gravitational collapse,
the charged matter fields which are left outside the horizon of the
newly born black hole would be radiated away to infinity or
eventually be absorbed into the black hole itself.
\bibitem{NoteQ} We shall assume, without loss of generality, that
$Q>0$.
\bibitem{Noterg} Note that the regime (\ref{Eq7}) corresponds to the
relation $3r_--r_+<0$, in which case one finds from (\ref{Eq6})
$\Im\omega_0>\pi T_{\text{BH}}$. This inequality violates the
universal relaxation bound (\ref{Eq2}) [or equivalently, it violates
the black-hole quasinormal bound (\ref{Eq3})].
\bibitem{Notehb} Note that the inequality (\ref{Eq2}) explicitly contains the
quantum Planck constant $\hbar$.
\bibitem{Notebq} Note that the inequality (\ref{Eq9}) contains an explicit
quantum (Planck constant) factor: $qQ\ll\mu^2r^2_+/\hbar$.
\bibitem{Noteqm} Note that $\mu$ and $q$ stand for $\mu/\hbar$ and $q/\hbar$, respectively.
Hence, these field parameters have the dimensions of
$($length$)^{-1}$.
\bibitem{Noteom} We shall henceforth omit the harmonic indices $l$ and $m$ for brevity.
\bibitem{Noteleq} The last inequality in (\ref{Eq27}) is a direct consequence of the strong inequality
(\ref{Eq23}) and the facts that $\tau\leq1$ [see Eq. (\ref{Eq25})]
and $H(r_+)\leq\mu^2r^2_++l(l+1)+1$ [see Eq. (\ref{Eq21})].
\bibitem{Noteie1} Note that the quantum constraint (\ref{Eq9}), together with the inequality
$H(r_+)\geq \mu^2r^2_+$ [see Eq. (\ref{Eq21})], imply $qQ/H(r_+)\leq
qQ/\mu^2r^2_+\ll1$.
\bibitem{Notemqq} Here we have used the relation [see Eq. (\ref{Eq21})]
$H(r_+)=\mu^2r^2_+\{1+O[l(l+1)/\mu^2r^2_+]\}$ in the regime $\mu
r_+\gg l+1$ [see Eq. (\ref{Eq41})].
\bibitem{Notecc} It is worth emphasizing again that Ref. \cite{Schw4} has
shown that the Schwinger-type pair production mechanism becomes
effective (that is, with pair production probability of almost
$100\%$) already at $E\simeq 0.03E_{\text{c}}\ll E_{\text{c}}$.
Thus, the quantum Schwinger discharge of charged black holes
enforces the strong inequality $E_+/E_{\text{c}}\ll1$.
\end{thebibliography}
\end{document}